%% file: TargetNet_Bioinformatics.tex
\documentclass{formatting/bioinfo}
\copyrightyear{2021} \pubyear{2021}
\access{Advance Access Publication Date: Day Month Year}
\appnotes{Manuscript Category}

\usepackage{amsmath,adjustbox,arydshln,booktabs,multirow,nameref,textgreek}
\usepackage{algorithm,algorithmicx,algpseudocode}
\usepackage[para]{threeparttable}
\usepackage{balance}

\newcommand{\eg}{{\it e.g.}}
\newcommand{\ie}{{\it i.e.}}

\begin{document}
\firstpage{1}

\subtitle{Subject Section}
\title[TargetNet: Functional microRNA Target Prediction with Deep Neural Networks]{TargetNet: Functional microRNA Target Prediction with Deep Neural Networks}
\input{0_1_authors.tex}
\corresp{$^\ast$To whom correspondence should be addressed.}
\editor{Associate Editor: XXXXXXX}
\history{Received on XXXXX; revised on XXXXX; accepted on XXXXX \vspace{0.3cm}}
\input{0_2_abstract.tex}
\maketitle

\section{Introduction}
\input{1_introduction.tex}

\section{Background}
\input{2_background.tex}

\section{Methods}
\input{3_1_methods_cts.tex}

\input{3_2_methods_input.tex}

\input{3_3_methods_resnet.tex}

\input{3_4_methods_post.tex}

\section{Experiments} 
\input{4_1_experiments_datasets.tex}

\input{4_2_experiments_test.tex}

\input{4_3_experiments_ablation.tex}

\section{Concluding Remarks}
\input{5_concluding_remarks.tex}

\section*{Acknowledgements}
\input{0_3_acknowledgment.tex}

\bibliographystyle{formatting/natbib}
\bibliography{references}

\end{document}

%% file: 0_1_authors.tex
\author[Min \textit{et~al}.]{Seonwoo Min\,$^{\text{\sfb 1,2}}$, Byunghan Lee\,$^{\text{\sfb 3}*}$, and Sungroh Yoon\,$^{\text{\sfb 1,4}*}$}
\address{$^{\text{\sf 1}}$Department of Electrical and Computer Engineering, Seoul National University, Seoul 08826, South Korea\\
$^{\text{\sf 2}}$LG AI Research, Seoul 07796, South Korea\\
$^{\text{\sf 3}}$Department of Electronic and IT Media Engineering, Seoul National University of Science and Technology, Seoul 01811, South Korea\\
$^{\text{\sf 4}}$Interdisciplinary Program in Artificial Intelligence and Interdisciplinary Program in Bioinformatics, Seoul National University, Seoul 08826, South Korea\\
}

%% file: 0_2_abstract.tex
\abstract{\textbf{Motivation:} 
MicroRNAs (miRNAs) play pivotal roles in gene expression regulation by binding to target sites of messenger RNAs (mRNAs). While identifying functional targets of miRNAs is of utmost importance, their prediction remains a great challenge. Previous computational algorithms have major limitations. They use conservative candidate target site (CTS) selection criteria mainly focusing on canonical site types, rely on laborious and time-consuming manual feature extraction, and do not fully capitalize on the information underlying miRNA-CTS interactions. \\
\textbf{Results:} 
In this paper, we introduce TargetNet, a novel deep learning-based algorithm for functional miRNA target prediction. To address the limitations of previous approaches, TargetNet has three key components: (1) relaxed CTS selection criteria accommodating irregularities in the seed region, (2) a novel miRNA-CTS sequence encoding scheme incorporating extended seed region alignments, and (3) a deep residual network-based prediction model. The proposed model was trained with miRNA-CTS pair datasets and evaluated with miRNA-mRNA pair datasets. TargetNet advances the previous state-of-the-art algorithms used in functional miRNA target classification. Furthermore, it demonstrates great potential for distinguishing high-functional miRNA targets. \\
\textbf{Availability:} The codes and pre-trained models are available at \href{https://github.com/mswzeus/TargetNet}{https://github.com/mswzeus/TargetNet}.\\	
\textbf{Contact:} B.L. (\href{bhlee@seoultech.ac.kr}{bhlee@seoultech.ac.kr}) or S.Y. (\href{sryoon@snu.ac.kr}{sryoon@snu.ac.kr})}

%% file: 1_introduction.tex
Gene expression regulation is a key component of biological processes. The expression levels of different genes are controlled through several mechanisms. MicroRNAs (miRNAs) play a pivotal role in the post-transcriptional regulation of $\geq$ 60\% of human protein-coding genes \citep{bartel2009micrornas}. MiRNAs are small non-coding RNAs that can bind to the target sites of messenger RNAs (mRNAs). This binding leads to the repression of efficient translation of mRNAs, thereby down-regulating the expression of target genes \citep{garcia2011weak}. The effectiveness of each target site can vary depending on the site context and the binding stability \citep{grimson2007microrna}. While identifying functional targets of miRNAs is of utmost importance, their computational prediction remains a great challenge \citep{kim2016general}. 

A miRNA can target multiple mRNAs by functioning as a sequence-specific guide. The binding is primarily directed through the interaction between the 5' ends of a miRNA, referred to as the "seed region," and the complementary 3' untranslated regions (UTRs) of a target mRNA. Previous large-scale transcriptome studies have identified several target canonical site types that form Watson-Crick (WC) parings with the miRNA seed region \citep{krek2005combinatorial}. The canonical site types include 6-mer sites (matching miRNA nucleotides 2-7), 7-mer-m8 sites (matching miRNA nucleotides 2-8), 7-mer-A1 sites (matching miRNA nucleotides 2-7 with an \texttt{A} opposite nucleotide 1), and 8-mer sites (matching miRNA nucleotides 2-8 with an \texttt{A} opposite nucleotide 1). More recent studies have also revealed that target non-canonical site types with \texttt{G}:\texttt{U} wobble pairings or gaps are also prevalent \citep{kim2016general, broughton2016pairing}. 

A variety of computational algorithms have been proposed for functional miRNA target prediction \citep{kern2019s}. Most of them follow a similar pipeline consisting of three stages. The first stage is the selection of candidate target sites (CTSs). Given a miRNA-mRNA pair, computational algorithms use a sliding window to identify CTSs from 3' UTRs of the mRNA fulfilling certain criteria. Then, in the second stage, a prediction model is used to identify whether each miRNA-CTS pair is functional or non-functional. Finally, in the third stage, the predictions are post-processed to obtain a final prediction for the miRNA-mRNA pair. In general, a miRNA-mRNA pair is predicted to be functional if there is at least one miRNA-CTS pair predicted as functional. 

While previous computational algorithms differ in CTS selection criteria and prediction models, they share certain major limitations. They generally use conservative CTS selection criteria which mainly focus on canonical site types. Because these conservative criteria only allow a limited number of non-canonical site types with few irregularities, they cannot capture the complete picture of functional miRNA target prediction \citep{kertesz2007role}. In addition, the majority of prediction models are based on feature extraction followed by application of conventional machine learning classifiers (\eg, linear regression and support vector machines). They rely on the discovery of new hand-crafted features, and often exploits additional information such as site location, accessibility, or minimum free energy \citep{kern2019s}. Nevertheless, manual feature extraction requires laborious and time-consuming processes. This inevitably impedes the improvement of prediction models in terms of both efficiency and performance \citep{min2017deep}.

Several studies have recently proposed deep learning-based prediction models \citep{lee2016deeptarget, pla2018miraw}. They eliminate manual feature extraction and use deep neural networks to automatically learn effective features. However, they still have not fully capitalized on information underlying miRNA-CTS interactions. Even though the CTS selection stage provides information on how each CTS forms pairings, mismatches, or gaps to bind with the miRNA seed region, previous studies only used miRNA-CTS sequences for their prediction models. This leaves considerable room for improvement and development of a more effective data-driven computational algorithm.

In this paper, we introduce TargetNet, a novel deep learning-based algorithm for functional miRNA target prediction  (Figure \ref{fig:schematic}). To address the previous limitations, TargetNet has three key components. First, TargetNet uses relaxed CTS selection criteria. Employing a sliding window, we align the extended seed region of a miRNA to the UTRs of a target mRNA. Then, we consider those aligned regions with at least 6 WC or wobble base pairings as the CTSs. Second, TargetNet uses a novel encoding scheme for miRNA-CTS sequences to incorporate extended seed alignment information. This makes it easier for the deep neural network to learn features from the bindings formed by a miRNA-CTS pair. Third, TargetNet uses a deep residual network (ResNet) with one-dimensional convolutions as its prediction model \citep{he2016deep}. Compared to previously used multilayer perceptrons and recurrent neural networks (RNNs), it can be more effective for RNAs where local nucleotide motifs often have significant implications. 

We used experimentally verified public datasets for empirical validations \citep{pla2018miraw, grimson2007microrna, paraskevopoulou2018microclip}. TargetNet was trained with miRNA-CTS pair datasets, and evaluated using miRNA-mRNA pair datasets. Leveraged by these three key components, TargetNet demonstrates significant performance improvement in functional miRNA target classification over previous state-of-the-art algorithms. Furthermore, top-ranked TargetNet prediction scores exhibit a considerable association with the level of miRNA-mRNA expression down-regulation, which demonstrates its great potential for distinguishing high-functional miRNA targets.
\vspace{0.5cm}

\input{figure/schematic}

%% file: figure/schematic.tex
\begin{figure}[t]
    \centering
	\includegraphics[width=0.76\columnwidth]{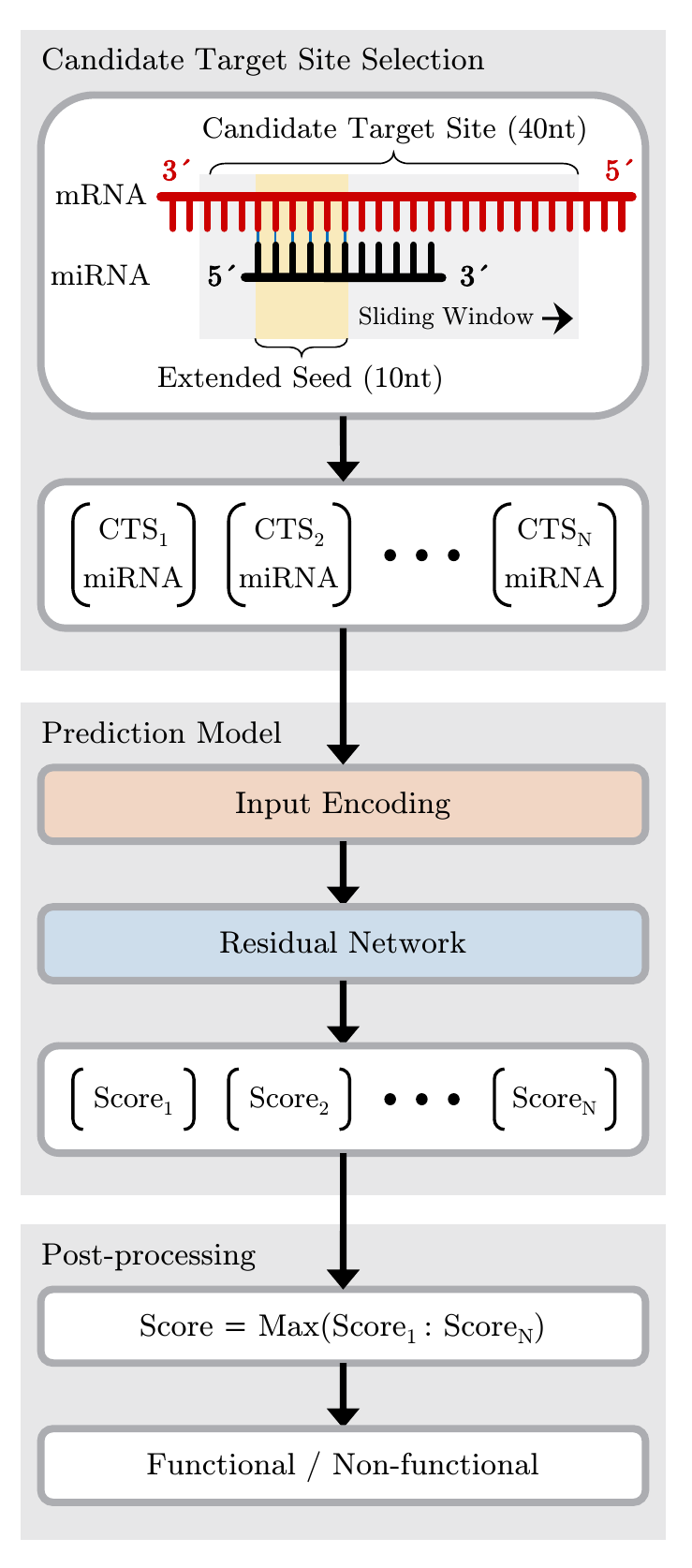}
   	\vspace{-0.5cm}
   	\caption{Schematic of TargetNet for functional microRNA target prediction.}
    \label{fig:schematic}
    \vspace{-0.5cm}
\end{figure}

%% file: 2_background.tex

\subsection{Candidate target site selection}
Since a miRNA partially forms WC pairings to its cognate target mRNAs, it is important to search for CTSs based on their binding characteristics in order to reduce the search space of a prediction algorithm.

In the literature, researchers search for CTSs that contain predefined site patterns as follows: \cite{kertesz2007role} considered (a) 7-mer of a mRNA that forms complete WC parings to a miRNA starting at nucleotide 1 or 2 and (b) a region that contains at least seven WC parings to a miRNA starting at nucleotide 1. \cite{agarwal2015predicting} considered (a) 6-mer of a mRNA that forms complete WC parings to a miRNA starting at nucleotide 2, (b) 4-mer of a mRNA that forms three consecutive complete WC parings to a miRNA starting at nucleotide 13, and (c) 12-mer of a mRNA that forms eleven consecutive complete WC parings to a miRNA starting at nucleotide 4. \cite{pla2018miraw} considered (a) a region that contains at least six WC parings to a miRNA at the nucleotides 1--10, (b) a region which contains at least seven WC parings to a miRNA at nucleotides 1--10, and (c) a region containing at least seven WC parings to a miRNA at nucleotides 2--10. 

They enable a target prediction algorithm to reduce false positives by pre-processing CTSs based on empirical observations; however, they cannot capture non-canonical site patterns~\citep{kim2016general}. To address this limitation, we used relaxed site patterns for searching CTSs.

\subsection{Prediction model}
In this paper, we categorized miRNA target prediction models into two types: feature extraction-based and deep learning-based models. For example, PITA~\citep{kertesz2007role}, mirSVR~\citep{betel2010comprehensive}, miRDB~\citep{wong2015mirdb}, and TargetScan~\citep{agarwal2015predicting} are feature extraction-based models, while deepTarget~\citep{lee2016deeptarget} and miRAW~\citep{pla2018miraw} are deep learning-based models.

\subsubsection{Feature extraction-based models}
It is known that $\geq$ 96 features contribute to miRNA target binding~\citep{liu2019prediction}. Existing feature extraction-based models utilize different features from each other. PITA utilizes site accessibility to compute a dynamic programming-based score. mirSVR utilizes sequence and contextual features to train a regression model. miRDB utilizes seed conservation features to train a support vector machine model. TargetScan utilizes seed conservation and structural features to train a regression model. Each hand-crafted feature engineering procedure depends on the research design; hence, it is difficult to define a consistent strategy.

\subsubsection{Deep learning-based models}
Some of the deep learning-based models use manually extract features as inputs for a model~\citep{cheng2015mirtdl}. In this study, we considered models that utilize raw sequences. deepTarget allows some non-canonical site types and utilizes RNN-based auto-encoders to learn features; however, it utilizes simulated training data to compensate for the number of negative pairs. miRAW utilizes multi-layer perceptron networks to learn features; however, it requires additional information, including binding and site accessibility energies. Although both models exploit CTSs to reduce the search space of their algorithms, they ignore the information underlying CTSs such as how each CTS forms pairings, mismatches, or bulges. To fully utilize the information underlying CTSs, we proposed a novel input encoding scheme for miRNA-mRNA pairs.

%% file: 3_1_methods_cts.tex
\subsection{Candidate target site selection}
\label{sec:cts}

Given a miRNA-mRNA pair, TargetNet first identifies CTSs that have the potential to be binding sites. We utilized a sliding window to scan through 3' UTRs of the mRNA (Figure \ref{fig:schematic}). Since nucleotides beyond the seed are also important for miRNA-CTS interaction \citep{sheu2019beyond}, we set the sliding window length to 40 nucleotides and its step length as 1 nucleotide. For each step, the sliding window produces a potential miRNA-CTS pair that is checked against the CTS selection criteria. In the following, miRNA and CTS sequences are denoted as:
\begin{equation*}
\begin{gathered}
    S^{\text{miRNA}} = (s^{\text{miRNA}}_{1}, \cdots, s^{\text{miRNA}}_{L_{\mathrm{mi}}}),\\
    S_{i}^{\text{CTS}} = (s^{\text{CTS}}_{i, 1}, \cdots, s^{\text{CTS}}_{i, 40}), \quad
    s^{\text{miRNA}}_{j}, s^{\text{CTS}}_{i, j} \in \{\texttt{A}, \texttt{U}, \texttt{G}, \texttt{C}\}, 
\end{gathered}
\end{equation*}
where $S^{\text{miRNA}}$ and $S_{i}^{\text{CTS}}$ are in the 5'-to-3' and 3'-to-5' directions, respectively. We use subscript $i$ to indicate multiple CTSs for a given miRNA-mRNA pair. $S^{\text{miRNA}}$ has a variable length, $L_{\mathrm{mi}}$, which is 22 nucleotides on average while $S_{i}^{\text{CTS}}$ has a fixed length of 40 nucleotides.

TargetNet adopts relaxed CTS selection criteria similar to those used in miRAW. First, we divide the $S^{\text{miRNA}}$ into sub-sequences as:
\begin{equation*}
\begin{gathered}
    S^{\text{miRNA}} = \langle S^{\text{miRNA-ES}}, S^{\text{miRNA-DS}} \rangle,\\ 
    S^{\text{miRNA-ES}} = (s^{\text{miRNA}}_{1}, \cdots, s^{\text{miRNA}}_{10}),\\
    S^{\text{miRNA-DS}} = (s^{\text{miRNA}}_{11}, \cdots, s^{\text{miRNA}}_{L_{\mathrm{mi}}}),\\
\end{gathered}
\end{equation*}
where $S^{\text{miRNA-ES}}$ and $S^{\text{miRNA-DS}}$ denote the extended seed region and downstream nucleotides of a miRNA sequence, respectively. Similarly, we divide $S^{\text{CTS}}$ into sub-sequences as:
\begin{equation*}
\begin{gathered}
    S_{i}^{\text{CTS}} = \langle S_{i}^{\text{CTS-DS}}, S_{i}^{\text{CTS-ES}}, S_{i}^{\text{CTS-US}} \rangle,\\
    S_{i}^{\text{CTS-DS}} = (s^{\text{CTS}}_{i, 1}, \cdots, s^{\text{CTS}}_{i, 5}),\\
    S_{i}^{\text{CTS-ES}} = (s^{\text{CTS}}_{i, 6}, \cdots, s^{\text{CTS}}_{i, 15}), \quad
    S_{i}^{\text{CTS-US}} = (s^{\text{CTS}}_{i, 16}, \cdots, s^{\text{CTS}}_{i, 40}),
\end{gathered}
\end{equation*}
where  $S_{i}^{\text{CTS-DS}}$, $S_{i}^{\text{CTS-ES}}$, and $S_{i}^{\text{CTS-US}}$ denote the downstream, extended seed region, and upstream nucleotides of a CTS sequence, respectively. Note that since $S_{i}^{\text{CTS}}$ is in a 3'-to-5' direction, the former sub-sequence is toward the 3' end, and hence, it is called $S_{i}^{\text{CTS-DS}}$. 

Then, we conduct a sequence alignment of the extended seed regions:
\begin{equation*}
    \widetilde{S}^{\text{miRNA-ES}}, \widetilde{S}_{i}^{\text{CTS-ES}} = \mathrm{Align}(S^{\text{miRNA-ES}}, S_{i}^{\text{CTS-ES}}). \\
\end{equation*}
We find their best global alignment using a Biopython pairwise2 package \citep{cock2009biopython}. The scoring matrix for the alignment is defined to produce a score of 1 for WC and wobble pairings, and a score of 0 for the other pairings and gaps. If there are multiple best alignments, we use the first one obtained from the package. The alignment results, $\widetilde{S}^{\text{miRNA-ES}}$ and $\widetilde{S}^{\text{CTS-ES}}$, are composed of $s \in \{\texttt{A}, \texttt{U}, \texttt{G}, \texttt{C}, \texttt{-}\}$ representing four nucleotides and a gap. The relaxed CTS selection criteria are met if the alignment score is at least 6. It makes minimal assumptions regarding miRNA-CTS interactions; hence, it can accommodate a wide range of non-canonical sites, as well as canonical sites. 

\input{figure/overview}

%% file: figure/overview.tex
\begin{figure*}[t]
    \centering
	\includegraphics[width=0.8\textwidth]{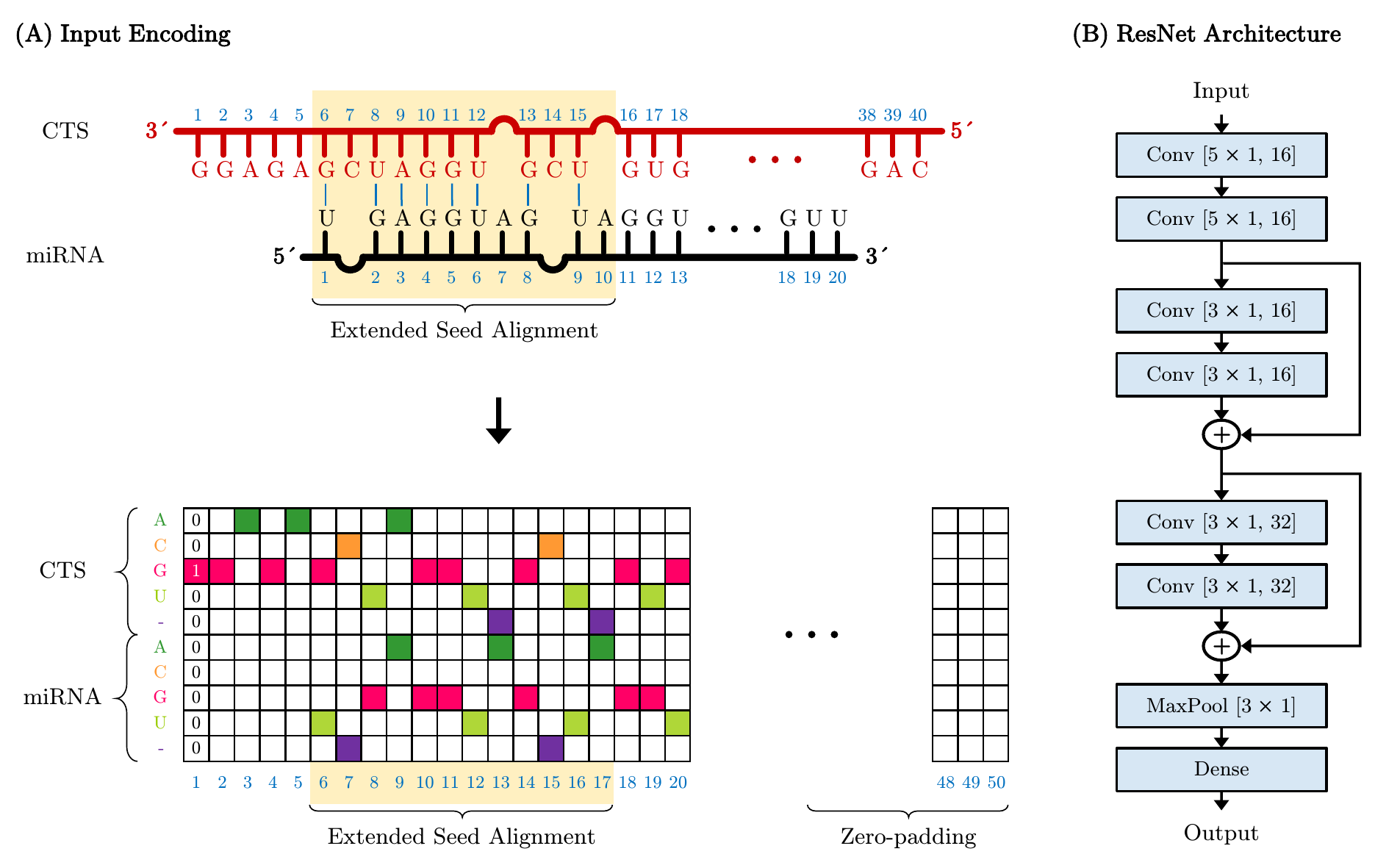}
   	\caption{Overview of TargetNet prediction model. (A) Input encoding. (B) ResNet architecture.}
    \label{fig:overview}
\end{figure*}

%% file: 3_2_methods_input.tex
\subsection{Input encoding}
\label{sec:input}
The most distinguishing component of TargetNet which separates it from other deep learning-based methods is the way it encodes a miRNA-CTS pair. Once the CTS selection is completed, the previous works use one-hot encoding to convert only sequences into numerical representations. In contrast, we propose a novel encoding scheme to incorporate additional information on how the extended seed regions of a miRNA-CTS pair are aligned and form bindings (Figure \ref{fig:overview}(A)). 

TargetNet input encoding takes the alignment results of the extended seed regions (\ie, $\widetilde{S}^{\text{miRNA-ES}}$ and $\widetilde{S}_{i}^{\text{CTS-ES}}$) in addition to the miRNA-CTS sequences (\ie, $S^{\text{miRNA}}$ and $S_{i}^{\text{CTS}}$). Specifically, we replace the extended region sequences, $S^{\text{miRNA-ES}}$ and $S_{i}^{\text{CTS-ES}}$, with their alignment results, $\widetilde{S}^{\text{miRNA-ES}}$ and $\widetilde{S}_{i}^{\text{CTS-ES}}$, and convert them using one-hot encoding:
\begin{equation*}
\begin{split}
\textbf{E}^{\text{miRNA}} & = \mathrm{Encode}(\langle \widetilde{S}^{\text{miRNA-ES}}, S^{\text{miRNA-DS}} \rangle)\\
& = \langle \textbf{e}^{\text{miRNA}}_{1}, \cdots, \textbf{e}^{\text{miRNA}}_{L_{\mathrm{m}}} \rangle,\\
\textbf{E}_{i}^{\text{CTS}} & = \mathrm{Encode}(\langle S_{i}^{\text{CTS-DS}}, \widetilde{S}_{i}^{\text{CTS-ES}}, S_{i}^{\text{CTS-US}} \rangle)\\
& = \langle \textbf{e}^{\text{CTS}}_{i, 1}, \cdots, \textbf{e}^{\text{CTS}}_{i, L_{\mathrm{c}}} \rangle,\\
\end{split}
\end{equation*}
where $\textbf{e}^{\text{miRNA}}_{j}$ and $\textbf{e}^{\text{CTS}}_{i, j}$ are 5-dimensional one-hot vectors indicating that the position is one of the 5 possible characters, $\{\texttt{A}, \texttt{U}, \texttt{G}, \texttt{C}, \texttt{-}\}$. Both $\textbf{E}^{\text{miRNA}}$ and $\textbf{E}_{i}^{\text{CTS}}$ have variable lengths (\ie, $L_{\mathrm{m}}$ and $L_{\mathrm{c}}$) due to possible gaps in the alignment results. 

Finally, we perform position-wise concatenation of $\textbf{E}^{\text{miRNA}}$ and $\textbf{E}_{i}^{\text{CTS}}$ with additional zero-padding $\textbf{0}^{\mathrm{k}} \in \mathbb{R}^{5 \times \mathrm{k}}$ as:
\begin{equation*}
\begin{gathered}
\textbf{E}_{i} = \mathrm{Concat}(\hat{\textbf{E}}^{\text{miRNA}}, \hat{\textbf{E}}_{i}^{\text{CTS}}),\\
\hat{\textbf{E}}^{\text{miRNA}} = \langle \textbf{0}^{5}, \textbf{E}^{\text{miRNA}}, \textbf{0}^{45-L_{\mathrm{m}}} \rangle, \quad 
\hat{\textbf{E}}_{i}^{\text{CTS}} = \langle \textbf{E}_{i}^{\text{CTS}}, \textbf{0}^{50-L_{\mathrm{c}}} \rangle.\\ 
\end{gathered}
\end{equation*}
Zero-paddings are used (1) to align the positions of the extended seed regions and (2) to make $\textbf{E}_{i}$ be a 10-by-50 sized vector. The advantage of the proposed input encoding is that it makes easier for the following ResNet to fully capitalize on information underlying miRNA-CTS interactions. The input vector is now able to represent not only the miRNA-CTS sequences but also how pairings, mismatches, or gaps are formed within their extended seed regions.

%% file: 3_3_methods_resnet.tex
\subsection{Residual Network}
\label{sec:resnet}
Motivated by the recent successes of ResNets in computer vision problems \citep{he2016deep}, we use a ResNet as a prediction model for TargetNet (Figure \ref{fig:overview}(B)). To the best of our knowledge, this is the first work to use the ResNet for functional miRNA target prediction.

Let $f_{k, n}$ be a one-dimensional convolution layer, where $k$ and $n$ denote the filter lengths and the number of filters, respectively. The filters are convolved along the length axis and learned to find motifs. Compared to other layers, it can be more effective for RNAs where motifs have significant implications. Furthermore, the filters can also be understood to be learnable position-weighted matrices used in conventional techniques \citep{min2017deep}. Each convolution layer is followed by a rectified linear unit (ReLU) activation function and dropout regularization \citep{srivastava2014dropout}. We use zero-paddings to keep the output sizes unchanged. 

First, our model has an input stem (denoted as $\mathrm{Stem}$) which takes the encoded miRNA-CTS vector as input:
\begin{equation*}
\begin{gathered}
\textbf{H}_{i, 1} = \mathrm{Stem}(\textbf{E}_{i}) = f_{5, 16}(f_{5, 16}(\textbf{E}_{i}, \textbf{W}_{1}), \textbf{W}_{2}),
\end{gathered}
\end{equation*}

Then, its output is feed into the two residual blocks (denoted as $\mathrm{ResBlock}_{1}$ and $\mathrm{ResBlock}_{2}$), each consisting of two convolution layers:
\begin{equation*}
\begin{gathered}
\textbf{H}_{i, 2} = \mathrm{ResBlock}_{1}(\textbf{H}_{i, 1}) = \textbf{H}_{i, 1} + f_{3, 16}(f_{3, 16}(\textbf{H}_{i, 1}, \textbf{W}_{3}), \textbf{W}_{4}),\\
\textbf{H}_{i, 3} = \mathrm{ResBlock}_{2}(\textbf{H}_{i, 2}) = \textbf{H}_{i, 2} + f_{3, 32}(f_{3, 32}(\textbf{H}_{i, 2}, \textbf{W}_{5}), \textbf{W}_{6}),
\end{gathered}
\end{equation*}
where $\textbf{W}_{l}$ is the learnable parameters of the $l$-th layer. Let $\mathcal{F}(\textbf{X})$ be an optimal function to be learned by a group of layers. While standard layers (\eg, input stem) are formulated to learn $\mathcal{F}(\textbf{X})$ directly, residual blocks are reformulated with skip connections to learn its residual function, $\mathcal{R}(\textbf{X}) := \mathcal{F}(\textbf{X}) - \textbf{X}$. It has been shown to ease learning by enabling us (1) to pre-condition $\mathcal{F}(\textbf{X})$ to be closer to an identity mapping and (2) to directly propagate forward and backward signals \citep{he2016identity}.

Finally, we compute the output score $0 \leq p_{i}^{\text{miRNA-CTS}} \leq 1$, which indicates how likely a given miRNA-CTS pair is functional. $\textbf{H}_{i, 3}$ is fed into a max-pooling layer and a dense layer (denoted as $\mathrm{MaxPool}$ and $\mathrm{Dense}$, respectively) as:
\begin{equation*}
p_{i}^{\text{miRNA-CTS}} = \mathrm{Dense}(\mathrm{MaxPool}(\textbf{H}_{i,3}), \textbf{W}_{7}),
\end{equation*}
where $\textbf{W}_{7}$ denotes the learnable parameters of the dense layer. The max-pooling layer reduces the output sizes by computing the channel-wise maximum value for non-overlapping windows of size 3. We use a sigmoid function as an activation function for the dense layer. 

For training of the ResNet prediction model, we use binary cross-entropy objective function defined as:
\begin{equation*}
{\mathcal{L}} = -(y\log(p^{\text{miRNA-CTS}}) + (1-y)\log(1-p^{\text{miRNA-CTS}})),
\end{equation*}
where $y \in \{0, 1\}$ specifies the label for a given miRNA-CTS pair. Note that we use miRNA-CTS pair datasets to train the prediction model rather than miRNA-mRNA pair datasets (Section \ref{sec:miRNA-CTS_dataset}). We use Adam optimizer \citep{kingma2014}, a training epoch size of 50, a mini-batch size of 256, a learning rate of 0.001, and a dropout probability of 0.5.

\input{table/results_miRAW_test}

%% file: table/results_miRAW_test.tex
\begin{table*}[t!]
    \centering
 	\caption{Comparison of functional miRNA target classification performance}
  	\label{table:result_miRAW_test}
  	\footnotesize
    \begin{threeparttable}
    \begin{tabular*}{\textwidth}{l@{\extracolsep{\fill}}cccccc}
        \toprule
            Method & F1 Score & Accuracy & Precision & Recall & Specificity & Negative Precision  \\  
        \midrule    
            PITA	 & 0.2162	 & 0.5053	 & 0.5196	 & 0.1365	 & 0.8741	 & 0.5030\\
            miRDB	 & 0.2110	 & 0.5373	 & 0.7135	 & 0.1239	 & \textbf{0.9507}	 & 0.5205\\
            miRanda	 & 0.3568	 & 0.5001	 & 0.4997	 & 0.2775	 & 0.7226	 & 0.5001\\
            TargetScan	 & 0.4712	 & 0.5577	 & 0.5852	 & 0.3945	 & 0.7208	 & 0.5436\\
        \addlinespace[0.6ex] \cdashline{1-7} \addlinespace[0.6ex]
            deepTarget	 & 0.4904	 & 0.6521	 & \textbf{0.8332}	 & 0.3477	 & 0.9354	 & 0.6064\\
            miRAW	  & 0.7289	 & 0.7055	 & 0.6749	 & 0.7923	 & 0.6186	 & 0.7493\\
        \addlinespace[0.6ex] \cdashline{1-7} \addlinespace[0.6ex]
            TargetNet (canonical) & 0.5205 & 0.6228 & 0.7136 & 0.4099 & 0.8358 & 0.5862 \\
            TargetNet (non-canonical) & 0.7736 & 0.7248 & 0.6570 & 0.9405 & 0.5091 & 0.8958 \\
            TargetNet (all) & \textbf{0.7739} & \textbf{0.7251} & 0.6572 & \textbf{0.9411} & 0.5091 & \textbf{0.8966} \\
        \bottomrule
    \end{tabular*}
    \end{threeparttable}
\end{table*}

%% file: 3_4_methods_post.tex
\subsection{Post-processing}
\label{sec:post}
In the final stage, the output scores are post-processed to obtain a final score for a miRNA-mRNA pair. We use the maximum value from the scores for each miRNA-CTS pair. Formally, if there are $N$ CTSs in a mRNA, we get output scores $p^{\text{miRNA-CTS}}_1, \cdots, p^{\text{miRNA-CTS}}_N$ for each CTS. The final score $p^{\text{miRNA-mRNA}}$ for a miRNA-mRNA pair is reported by:
\begin{equation*}
    p^{\text{miRNA-mRNA}} = \max(p^{\text{miRNA-CTS}}_1, \cdots, p^{\text{miRNA-CTS}}_N).
\end{equation*}
This results in the prediction of a miRNA-mRNA pair as functional if there is at least one functional miRNA-CTS pair. For the binary classification of functional targets, we used a threshold of 0.5 to binarize the final score $p^{\text{miRNA-mRNA}}$. Note that in contrast to miRAW, we do not exploit any additional filters based on site accessibility or minimum free energy.

%% file: 4_1_experiments_datasets.tex
\subsection{Datasets}
\subsubsection{miRNA-mRNA pair datasets}
\label{sec:miRNA-mRNA_dataset}
The complete TargetNet algorithm was evaluated with two types of experimentally verified miRNA-mRNA pair datasets, (1) miRAW and (2) log fold change (LFC) test datasets. 

First, we used miRAW test datasets with binary labels indicating functional and non-functional targets \citep{pla2018miraw}. They originated from DIANA-TarBase \citep{vlachos2015diana} and MirTarBase \citep{chou2016mirtarbase} databases. After removing duplicated samples, they consisted of 309,912 positive and 1,096 negative miRNA-mRNA pairs. Then, they were split in half and used for the train-validation (Section \ref{sec:miRNA-CTS_dataset}) and test datasets, respectively. The authors generated ten randomly sampled test datasets, each of which consisted of 548 positive and 548 negative pairs. The miRAW test datasets can help us evaluate the functional miRNA target classification performance of TargetNet. 

Second, we used eleven microarray and two RNA-seq LFC test datasets with real-valued labels indicating the level of functionality of miRNA targets \citep{grimson2007microrna, paraskevopoulou2018microclip}. In each microarray and RNA-seq dataset, a miRNA was individually transfected into HeLa and HEK329 cells, respectively. Then, the log fold change of mRNA expression was measured. More negative labels indicate more functional miRNA-mRNA pairs, which strongly down-regulate the targeted genes. The LFC test datasets can help us to evaluate how well TargetNet distinguishes high-functional miRNA targets.
\vspace{-0.2cm}

\subsubsection{miRNA-CTS pair datasets}
\label{sec:miRNA-CTS_dataset}
The prediction model of TargetNet was trained with miRAW miRNA-CTS pair datasets. To obtain miRNA-CTS pairs from the excluded miRNA-mRNA pairs, the authors pre-processed the positive pairs in two ways. One was cross-referencing with binding sites from PAR-CLIP \citep{grosswendt2014unambiguous} and CLASH \citep{helwak2013mapping}, and keeping miRNA-CTS pairs that form stable duplexes. The other was cross-referencing with conserved sites from TargetScanHuman \citep{agarwal2015predicting}. The negative pairs were pre-processed using a sliding window to identify miRNA-CTS pairs that also form stable duplexes. The stability was measured with RNACofold \citep{lorenz2011viennarna} by checking whether their secondary structures produce negative free energy.

Since the miRAW dataset split was based on miRNA-mRNA pairs (Section \ref{sec:miRNA-mRNA_dataset}), similar miRNAs can be distributed in both train-validation and test datasets. Thus, we intended to use independent LFC test datasets to further evaluate generalization performance in terms of different miRNAs. For this end, we filtered out miRNA-CTS pairs so that no two miRNAs from the miRAW train-validation and LFC test datasets have Levenshtein edit distance lower than 7. Then, we randomly selected 20 miRNAs and used corresponding 2,385 positive and 2,264 negative miRNA-CTS pairs as a validation set. The remaining pairs containing 26,803 positive and 27,341 negative pairs were used as a training set.

\input{table/results_miRAW_CTS}

%% file: table/results_miRAW_CTS.tex
\begin{table}[t!]
    \centering
 	\caption{miRNA target classification results with different CTS selection criteria}
  	\label{table:result_miRAW_CTS}
  	\footnotesize
    \begin{threeparttable}
    \begin{tabular*}{\columnwidth}{l@{\extracolsep{\fill}}lccc}
        \toprule
            Criteria & Method & F1 Score & Precision & Recall \\  
        \midrule    
            \multirow{2}{*}{miRAW-6-1:10}   & miRAW & 0.7289	& 0.6749	& 0.7923\\
            & TargetNet & 0.7491 & 0.7277 & 0.6945 \\
        \addlinespace[0.6ex] \cdashline{1-5} \addlinespace[0.6ex]
            \multirow{2}{*}{miRAW-7-1:10}   & miRAW & 0.7069	& 0.7188	& 0.6956\\
            & TargetNet & 0.7388 & 0.7242 & 0.7014 \\
        \addlinespace[0.6ex] \cdashline{1-5} \addlinespace[0.6ex]
            \multirow{2}{*}{miRAW-7-2:10}   & miRAW & 0.7222	& 0.7193	& 0.7255\\
            & TargetNet & 0.7422 & 0.7282 & 0.7056 \\
        \addlinespace[0.6ex] \cdashline{1-5} \addlinespace[0.6ex]
            \multirow{2}{*}{TargetScan}     & miRAW & 0.5325	& 0.7859	& 0.4029\\
            & TargetNet & 0.6747 & 0.6923 & 0.7155 \\
        \addlinespace[0.6ex] \cdashline{1-5} \addlinespace[0.6ex]
            \multirow{2}{*}{PITA}           & miRAW & 0.5694	& 0.7654	& 0.4537\\
            & TargetNet & 0.6901 & 0.6979 & 0.7081 \\
        \bottomrule
    \end{tabular*}
    \end{threeparttable}
\end{table}

%% file: 4_2_experiments_test.tex
\input{figure/results_LFC}

\subsection{Test results on miRNA-mRNA pair datasets}
\subsubsection{Classification of functional and non-functional targets}
\label{sec:experiments_classification}
First, we compared functional miRNA target classification performance of six different prediction algorithms: PITA \citep{kertesz2007role}, miRDB \citep{wong2015mirdb}, miRanda \citep{betel2010comprehensive}, TargetScan \citep{agarwal2015predicting}, deepTarget \citep{lee2016deeptarget} and miRAW \citep{pla2018miraw}. We excerpted the results of compared methods from the previous work \citep{pla2018miraw}, where the optimal reported configuration for each algorithm was employed.

Table \ref{table:result_miRAW_test} presents the averaged classification performance of ten test datasets. We used the miRAW test datasets, each of which consisted of 548 positive and 548 negative pairs (Section \ref{sec:miRNA-mRNA_dataset}). The results demonstrated that TargetNet considering both canonical and non-canonical sites outperforms the other state-of-the-art algorithms in terms of general performance measures, namely, F1 score and accuracy. The F1 score and accuracy differences between TargetNet and the second best algorithm, miRAW, were statistically significant with $p$-values of $1.1 \times 10^{-5}$ and $2.1 \times 10^{-3}$, respectively. For the evaluation of statistical significance, we used the two-sample Kolmogorov-Smirnov test \citep{massey1951kolmogorov}. Its rejected null hypothesis is that the two independent groups of samples (\eg, F1 scores obtained from TargetNet and miRAW) are from the same distributions.

While the other prediction models, PITA, miRDB, miRanda, TargetScan, and deepTarget, exhibited high specificity, they failed to correctly classify a large number of functional miRNA targets. This is largely due to their conservative CTS selection criteria, which neglect the majority of non-canonical site types. By comparing the performance of TargetNet for canonical and non-canonical sites, we could observe that non-canonical sites resulted in a significantly higher recall. It once again showed that accommodating non-canonical sites is vital for the classification of functional targets. Since TargetNet and miRAW share similar CTS selection criteria and the same training dataset, their comparison can illustrate the effectiveness of the prediction model. The performance improvement indicates that the proposed input encoding scheme and ResNet architecture can better capture the information underlying miRNA-CTS interactions. 

We investigated more closely how the CTS selection affect classification performance on miRAW test datasets. First, we compared the performance of TargetNet with and without the CTS selection. While they provided similar F1 scores (0.7739 vs. 0.7752), the latter took twice more time due to the additional non-filtered pairs. The results showed that the CTS selection is not required, but it can accelerate TargetNet by eliminating numerous negative pairs. Then, while keeping the other stages intact, we evaluated miRAW and TargetNet using five different CTS selection criteria (Table \ref{table:result_miRAW_CTS}). Note that miRAW-6-1:10 is identical to the one used in TargetNet, except that it uses a sliding window step length of 5 nucleotides. We can make the following observations from the results. First, regardless of the CTS selection criteria, TargetNet consistently outperformed miRAW in terms of the F1 score. This once again demonstrated the effectiveness of the proposed model. Second, using more conservative criteria generally deteriorates the classification performance of both miRAW and TargetNet. It filters out most of the candidate targets before using the prediction models, thus resulting in significant drops of recall and F1 score.

\subsubsection{Distinguishing high-functional targets}
Next, we examined the association between the level of expression down-regulation and the top-ranked prediction scores. We used independent eleven microarray and two RNA-seq LFC datasets that do not contain any miRNAs similar to those in the miRAW train-validation dataset. We compared TargetNet with two sets of state-of-the-art algorithms. For the microarray datasets, we compared with miRDB, TargetScan, PITA,  miRanda, deepTarget, and miRAW. We used publicly available codes to obtain prediction results for each algorithm. If an algorithm produces scores for each miRNA-CTS pair, we used the maximum value for summarizing them into scores for each miRNA-mRNA pair. For the RNA-seq datasets, we compared with microCLIP \citep{paraskevopoulou2018microclip}, TargetScan, MiIRZA \citep{khorshid2013biophysical}, PARma \citep{erhard2013parma}, and microMUMMIE with posterior and Viterbi decoding \citep{majoros2013microrna}. We excerpted the results of compared methods from the previous work \citep{paraskevopoulou2018microclip}. 

\input{table/results_miRAW_ablation}

Figure \ref{fig:results_lfc} presents the mean expression log fold changes of the top-ranked targets for microarray (A--B) and RNA-seq (C--D) datasets. We ranked each miRNA-mRNA pair from the LFC datasets according to the target prediction scores of each algorithm. Then, from the top-ranked target predictions, the averages of their expression log fold change values were plotted over a broad range of thresholds. First, we compared the performance of TargetNet for canonical and non-canonical sites allowed in the relaxed CTS selection criteria (Figure \ref{fig:results_lfc} (B) and (D)). While accommodating the non-canonical sites was vital in the classification of functional targets, it showed negative impacts for distinguishing high-functional targets. The results demonstrated that canonical sites with extensive seed binding are more crucial for the high-functional targets \citep{agarwal2015predicting}. Note that although the functionality of non-canonical sites is usually much weaker than those of canonical sites, they have been repeatedly validated through independent wet-lab analyses \citep{kim2016general}. It suggests that non-canonical sites are also likely to have biological roles and classifying all the functional targets is still important (Section \ref{sec:experiments_classification}). Next, we compared TargetNet for canonical sites with other state-of-the-art algorithms ((Figure \ref{fig:results_lfc} (A) and (C)). In both microarray and RNA-seq datasets, top TargetNet predictions were considerably associated with the level of target expression down-regulation. As we select a smaller number of top-ranked predictions from TargetNet, we can observe more repressed, thus, more functional targets. Even though the proposed algorithm does not exploit any expression training data, it shows comparable performance compared to the other state-of-the-art algorithms trained with expression data.

%% file: figure/results_LFC.tex
\begin{figure*}[ht!]
    \centering
	\includegraphics[width=\textwidth]{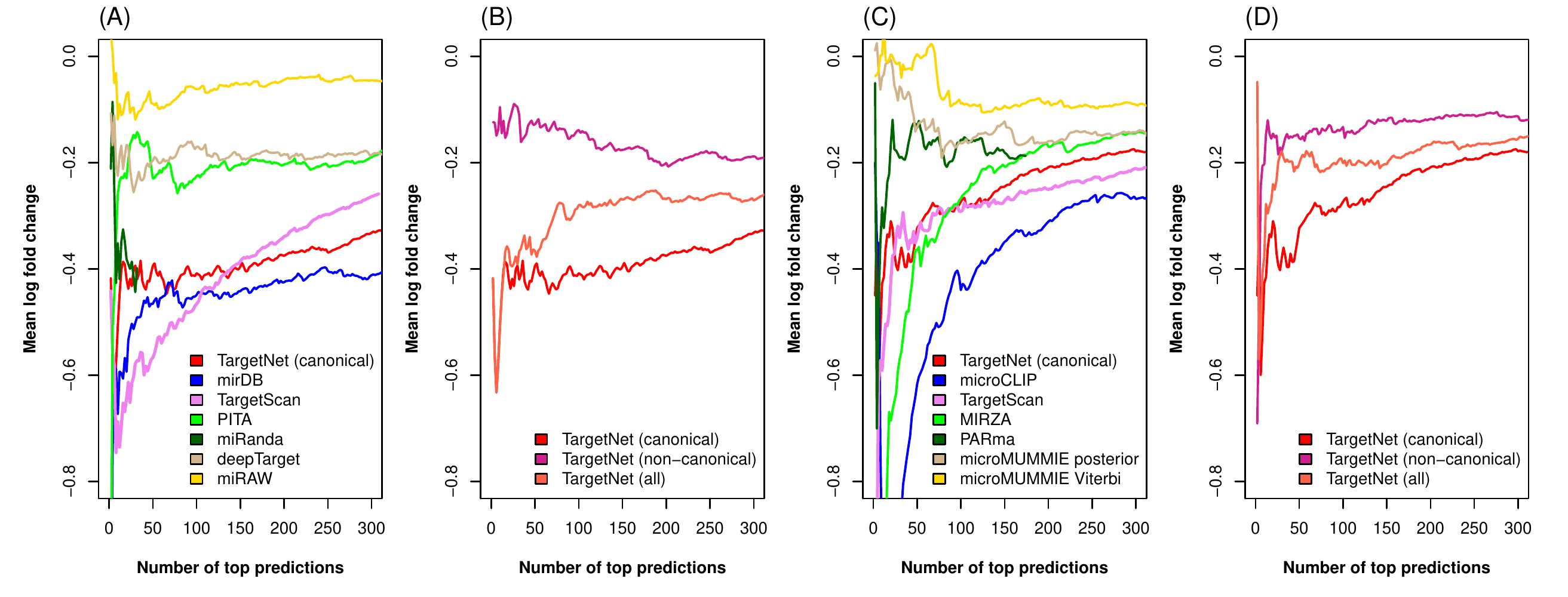}
   	\caption{Performance of prediction algorithms on distinguishing high-functional targets. Mean log fold changes of miRNA-mRNA expression down-regulation of the top-ranked targets were plotted. We used unified sets of eleven microarray (A--B) and two RNA-seq LFC (C--D)vudt datasets, in which a miRNA was individually transfected into HeLa and HEK329 cells, respectively. (A) and (C) compare TargetNet with other state-of-the-art algorithms. (B) and (D) show the effect of CTS selection for TargetNet in distinguishing high-functional targets.}
    \label{fig:results_lfc}
\end{figure*}

%% file: table/results_miRAW_ablation.tex
\begin{table}[t!]
    \centering
 	\caption{Ablation studies on the prediction model of TargetNet}
  	\label{table:result_miRAW_ablation}
  	\footnotesize
    \begin{threeparttable}
    \begin{tabular*}{\columnwidth}{c@{\extracolsep{\fill}}cccccc}
        \toprule
            & \shortstack{Alignment\\Encoding} & \shortstack{Skip\\Connection} & \shortstack{Number of\\Blocks} & \shortstack{Widening\\Factor} & F1 Score \\  
        \midrule    
            \textbf{BASE} & \textbf{TRUE} & \textbf{TRUE} & \textbf{2} & \textbf{1} & \textbf{0.8230}\\
        \addlinespace[0.6ex] \cdashline{1-6} \addlinespace[0.6ex]
            (A) & FALSE & TRUE & 2 & 1 & 0.7204\\
        \addlinespace[0.6ex] \cdashline{1-6} \addlinespace[0.6ex]
            (B) & TRUE & FALSE & 2 & 1 & 0.7743\\
        \addlinespace[0.6ex] \cdashline{1-6} \addlinespace[0.6ex]
            \multirow{3}{*}{(C)} & TRUE & TRUE & 2 & 0.5 & 0.7955\\
            & TRUE & TRUE & 2 & 2   & 0.8102\\
            & TRUE & TRUE & 4 & 1   & 0.7362\\
        \bottomrule
    \end{tabular*}
    \end{threeparttable}
    \vspace{-0.4cm}
\end{table}

%% file: 4_3_experiments_ablation.tex
\subsection{Ablation Studies}
Table \ref{table:result_miRAW_ablation} presents the results of the ablation studies to better understand TargetNet prediction models. We varied the components of the base model and measured the classification performance on the miRAW validation set. 

In row (A), we can observe that disregarding the proposed alignment input encoding significantly degrades the model performance. This suggests that incorporating extended seed region alignments provides invaluable information for functional miRNA prediction. In row (B), we replaced our ResNet model with a conventional convolutional neural network by removing the skip connections. Note that the compared model has the same number of parameters as the base model. Thus, the performance drop confirms that the residual connection enables more efficient training of the model. Finally, in rows (C), we varied the number of blocks and the number of filters by a widening factor. While doubling the number of filters produced similar results to the base model, other model complexity alterations resulted in inferior classification performance.

%% file: 5_concluding_remarks.tex
We proposed a deep learning-based algorithm for functional miRNA target prediction. TargetNet adopts relaxed CTS selection criteria to accommodate a variety of non-canonical and canonical site types. We introduced a novel input encoding scheme to embrace both miRNA-CTS sequences and how their extended seed regions form bindings. Then, we used ResNet to capture the information underlying miRNA-CTS interactions. Our experimental results supported that TargetNet not only demonstrates significant performance improvements in functional miRNA target classification, but also its top-ranked prediction scores show a considerable association with the level of miRNA-mRNA expression down-regulation.

%% file: 0_3_acknowledgment.tex
This work was supported by the National Research Foundation of Korea grants funded by the Ministry of Science and ICT (2018R1A2B3001628 (S.Y.), 2014M3C9A3063541 (S.Y.), 2019R1G1A1003253 (B.L.)), Institute of Information \& communications Technology Planning \& Evaluation (IITP) [NO.2021-0-01343, AI Graduate School Program] (S.Y.), the Ministry of Agriculture, Food and Rural Affairs (918013-4 (S.Y.)), and the Brain Korea 21 Plus Project in 2021 (S.Y.).